\documentclass[12pt]{article}

\usepackage{latexsym,amsmath,amssymb,epsfig}
\def\be{\begin{equation}}
\def\ee{\end{equation}}

\def\bea{\begin{eqnarray}}
\def\eea{\end{eqnarray}}
\def\nn{\nonumber \\}

\def\ie{{i.e.\ }}

\def\eps{{\epsilon}}

\def\bpsi{\bar\psi}
\def\bchi{\bar\chi}

\def\half{\frac{1}{2}}

\def\hc{{\rm h.c.}}

\def\Lag{{\mathcal{L}}}
\renewcommand{\O}{{\mathcal{O}}}

\def\d{\partial}
\def\dsl{{\not \! \partial}}

\def\R{{\mathrm{R}}}
\def\L{{\mathrm{L}}}

\newcommand{\refeq}[1]{\mbox{(\ref{#1})}}
\newcommand{\ltsim}{\lower3pt\hbox{$\, \buildrel < \over \sim \, $}}
\newcommand{\gtsim}{\lower3pt\hbox{$\, \buildrel > \over \sim \, $}}

\makeatletter 
\def\secteqno{\@addtoreset{equation}{section}%
\def\theequation{\thesection.\arabic{equation}}}

\begin{document}
\secteqno
\baselineskip=.58cm

\noindent

\begin{titlepage}
\begin{flushright}
{ ~}\vskip -1in
CERN-PH-TH/2006-011 \\
UG-FT-201/06 \\
CAFPE-71/06 \\
FERMILAB-PUB-06-012-T \\
hep-ph/0601222\\
January 2006\\
\end{flushright}

\vspace*{20pt}
\bigskip

\centerline{\Large  Effective description of brane terms in extra
  dimensions} 
\bigskip
\vskip 0.9truecm
\centerline{\sc
F.\ del Aguila$^{\,a}$,
M.\ P\'erez-Victoria$^{\,b}$,
and J.\ Santiago$^{\,c}$}

\vspace{1pc}

\begin{center}
{\em
 $^a$ CAFPE and Departamento de F\'\i sica Te\'orica y del Cosmos,\\
  Universidad de Granada, E-18071 Spain\\
\bigskip
$^b$ CERN, Theory Division, CH-1211 Geneva 23, Switzerland\\
\bigskip
$^c$ Fermi National Accelerator Laboratory, P.O.\ Box 500, Batavia,\\
IL 60510, USA} \\

\vspace{5pc}

{\large \bf Abstract}

\end{center}

We study how theories defined in (extra-dimensional) spaces with
localized defects can be 
described perturbatively by effective field  
theories in which the width of the defects vanishes. These effective 
theories must incorporate a ``classical'' renormalization, and we 
propose a renormalization prescription  {\it \`a
  la\/} dimensional regularization for codimension 1, which can be
easily used in phenomenological 
applications. As a check of the validity of this setting,
we compare some
general predictions of the renormalized effective theory with those
obtained in a particular ultraviolet completion based on
deconstruction. 

\end{titlepage}


\section{Introduction}

Many physical systems are described by fields propagating in a
space with lower--dimensional defects, including, in particular,
boundaries. These infinitely thin defects are typically idealizations
of localized physical backgrounds with finite size and a certain
substructure. The field theory should then be regarded as an effective
theory valid at low energies, such that the substructure of
the defects is not resolved. An implicit assumption underlying the
simplification of using 
zero-width (``thin'') defects is
that at low energies all observables are fairly insensitive to
ultraviolet details. As we will see, this is not always the
case: there are examples in which the details of the defects do not
decouple but filter into the low-energy observables. Nevertheless, we
will argue that in all cases an effective field theory with thin
defects can describe low-energy physics to any required precision. The
only difference between the decoupling and the non-decoupling
scenarios is that in the second case (part of) the substructure of the
brane is described by relevant operators.

On the other hand, it turns out that perturbative calculations
in the presence of thin defects are often plagued with
extra divergences that arise in the limit of zero thickness. In
some cases they appear already at the classical
level. These divergences signal a 
breakdown of the field theory at scales where the finite thickness of
the defects cannot be neglected. They must be renormalized away, and
the information about the microscopic structure of the defects is then
encoded in the renormalized coefficients of the different operators of
the theory. 

In this paper we propose a simple renormalization
prescription to deal with the divergences associated to thin defects
(thin-brane divergences), and study the effect of
bulk and localized higher-order terms. It turns out that, in our
scheme, the most singular (orthogonal) localized kinetic terms
can be completely eliminated via field redefinitions. This justifies the
conventional phenomenological approach of ignoring them. 
Furthermore, we check that the results obtained in the effective
framework agree with those given by a 
particular (deconstructed) microscopic 
theory. The latter will be described in detail in a
forthcoming publication~\cite{decorbs}. 

The paper is organized as
follows. In Section~2, we argue that effective theories
with infinitely thin defects are a good description of more
fundamental theories, in which the defects may have some internal
structure. We also introduce the particular setup to be studied in the
following sections. In Section~3, we describe our renormalization
scheme and show how it can be used to eliminate some operators to all
orders. In Section~4 we calculate, to second order in the derivative
expansion, the spectrum and wave functions of fermions, scalars and
gauge bosons for a general effective theory. 
Section~5 contains the matching to the
deconstructed models. Section~6 is devoted to a particular class of
operators, which are ambiguous in the limit of zero brane width. 
Finally, we present our conclusions in Section~7. 


\section{Effective theories with thin branes}

In the absence of defects, and under very general assumptions, any  
fundamental theory can be described, at energies below some scale
$\Lambda_0$, by an effective quantum field theory for the light degrees
of freedom~\cite{Weinberg:1978kz,Leutwyler:1993iq}. 
This $\Lambda_0$ is related to some characteristic
dimensionful parameter of the fundamental theory; it can also
represent a scale at which the effective theory becomes strongly
coupled. The subscript 0 is used to distinguish this scale from the
cutoff of the effective theory with defects. 
The effective Lagrangian can be 
expanded in an infinite series of local operators, organized in powers
of $E/\Lambda_0$ and  
$m_i/\Lambda_0$, where $E$ is the energy and $m_i$ represent possible
mass scales in the theory, smaller than
$\Lambda_0$. Of course, at
energies below a given $m_i$ one could describe physics by a new
effective theory with $\Lambda_0^\prime=m_i$. 
In a Wilsonian
framework, the scale $\Lambda_0$ represents some cutoff of
external and virtual momenta, such that the effective theory cannot
resolve distances smaller than $\Lambda_0^{-1}$. In practice, however,
it is more convenient to work with effective theories renormalized in
a mass-independent scheme, such as dimensional regularization with
minimal subtraction. The main
reason is that this prevents divergent loop corrections from enhancing
the effect of higher-order operators,
so that operators of order greater than a given $n$ do not contribute
to observables to order $\Lambda_0^{-n}$. Another reason is that
preserving symmetries in a Wilsonian context is more involved. In a
mass-independent scheme, the scale $\Lambda_0$ only appears in the
effective theory in the explicit inverse powers in front of the
different operators.

Consider now a generic ``fundamental'' theory defined in some
flat space that contains defects extended in $D$ infinite space-time
dimensions, with a characteristic thickness $\eps_0$, much smaller
than the size of the transverse dimensions $L$.  Furthermore, 
we restrict to plane defects with vanishing extrinsic curvature. These
requirements are not essential but 
simplify the discussion. The defects can 
have different miscroscopic origins: surfaces of materials, 
solitonic configurations, orientifold planes, D-branes,
intersections of D-branes, etc. In the following we will
generically call these objects ``branes''.
For simplicity, we assume that all other 
scales characterizing the brane (such as possible dimensionful
couplings between a localized background and the fluctuating fields)
are of the order of $\eps_0^{-1}$ as well. 
Because Poincar\'e invariance in the
transverse directions is broken, the corresponding
momenta are no longer good quantum numbers. It is then convenient
to work in position space for the transverse coordinates, and to
speak of derivatives, rather
than energies. D-momenta and Kaluza--Klein (KK) masses can also
label the eigenstates of the free Hamiltonian, but they are not
adequate  to organize local operators. We assume that, without branes,
an effective description exists in which the cutoff $\Lambda_0$ is
larger than the compactification scale $L^{-1}$. The latter can then
be thought of as one of the low scales $m_i$. 

When the branes are
introduced, we can distinguish two physical situations, 
according to the relative sizes of $\Lambda_0$ and $\eps_0^{-1}$.
If $\eps_0^{-1} < \Lambda_0$, it is
possible in principle to describe physics at energies below
$\Lambda_0$ by an 
effective field theory incorporating a field-theoretical
representation of the branes at scales between
$\eps_0^{-1}$ and $\Lambda_0$. One example of this situation is the
calculation of zero-point (Casimir)
energies of quantum electromagnetic fields in a conducting cavity or
in the presence of conducting plates, and toy models related to
this situation\footnote{See~\cite{Aghababaie:2003iw} for a discussion
  of effective theories and matching in the context of the Casimir
  effect, and~\cite{Graham:2003ib} for calculations with ``fat''
  branes. Note 
  that the localized energy density is a relevant operator in the
  effective theory below $\Lambda=\eps_0^{-1}$ (with thin branes
  represented by boundary conditions). Its coefficient, which is
  naturally controlled by the scale $\eps_0^{-1}$, is an input
  parameter 
  to be fixed by experiment or by matching with the theory with finite 
  $\eps_0$.}. 
If, instead, $\eps_0^{-1} \gtrsim \Lambda_0$
then the 
microscopic structure of the branes lies beyond the reach of the
effective field theory, and can only be
described at the level of the fundamental theory. This
situation is implicit in many field-theoretical models in extra
dimensions. In the example of fundamental theory we will consider
below, deconstruction, the scale  
$\Lambda_0$ (which is identified with the inverse of the lattice
spacing) acts as a hard cutoff in position space, which smears the 
brane over an effective size $\Lambda_0^{-1}$.

In all cases the theory can be described at
energies lower than $\Lambda=\mbox{min}\{\Lambda_0,\eps_0^{-1}\}$ by
an effective field theory with cutoff $\Lambda$~\footnote{When
  $\eps_0^{-1} < \Lambda_0$ we could use separate cutoffs for brane
  and bulk operators:
  $\Lambda_{\mathrm{brane}}=\eps_0^{-1}$,
  $\Lambda_{\mathrm{bulk}}=\Lambda_0$. This is analogous to the
  position-dependent cutoff 
  that is used in warped geometries. 
  Locality of the ultraviolet divergences implies that
  coefficients of brane operators do not appear in the running of
  the coefficients of bulk operators. The converse does not
  necessarily hold, but
  the suppressions by powers of $1/\Lambda_{\mathrm{brane}}$ are not
  destabilized by the addition of powers of
  $1/\Lambda_{\mathrm{bulk}}$. 
  Nevertheless we stick to the effective theory
  with a single cutoff $\Lambda$, although additional suppressions by
  powers of $1/(\eps_0 \Lambda_0)$ could be expected for dimensionless
  bulk couplings.}.
Note that if we send the physical scale $\eps_0^{-1}$ to infinity
the branes do not disappear, but become infinitely
thin. Correspondingly, the effective theory
lives in a space with ``effective'' branes of size
$\eps\leq \Lambda^{-1}<L$.  We stress that the auxiliary thickness
$\eps$ of the branes in the effective theory is not necessarily
related to the physical thickness $\eps_0$ of the branes in the
fundamental theory (which in some scenarios is represented by
$\Lambda$). Actually, at the end of the day we will send
$\eps\rightarrow 0$ to describe theories with finite~$\eps_0$. 

One important
feature of theories with branes is the appearance of localized
divergent radiative corrections, which implies
that brane localized terms must be included in the theory for
multiplicative
renormalizability~\cite{Symanzik:1981wd,McAvity:1992fq,Georgi:2000ks}.
In other words, putting 
them to zero is not stable under renormalization group
evolution. These brane operators can also be present at the scale
$\Lambda$, for instance if they are radiatively generated by heavy
degrees of freedom which have been integrated out. Operators with the
same field content 
are organized according to their canonical dimension (\ie the number
of derivatives and delta functions). This corresponds to an expansion
in $1/\Lambda$. 
In the Wilsonian framework, because 
the cutoff makes the theory
insensitive to distances smaller than $\Lambda^{-1}$,
we can use effective branes of any shape and size $\eps$ we like, as
long as the
integrated features of the original branes over a region of size
$~\Lambda^{-1}$ are preserved. If we knew the fundamental theory, the
process of integrating out the degrees of freedom higher than
$\Lambda$ would naturally give $\eps \approx \Lambda^{-1}$. On the
other hand, in a mass-independent scheme the theory will be sensitive
to the auxiliary scale $\eps^{-1}$. In this framework we
would like to keep $\Lambda$ as the only dimensionful scale (outside
loop logarithms), and eliminate the auxiliary brane thickness by
taking the thin-brane limit $\eps \rightarrow 0$. However, this is not
straightforward, for this limit can be divergent (after subtraction
of the usual 
divergences of correlators at coincident points). The divergences
arise because in the presence of certain brane terms, the fields (even 
for the lowest KK modes) fluctuate very strongly near the branes,
in such a way that  
the local value of the derivatives of the fields is of order
$\eps^{-1}$. Hence, perturbativity in $\d/\Lambda$ is spoiled.
The solution is,
as usual in field theory, to apply a renormalization procedure to
eliminate the dependence on $\eps$. The limit $\eps\rightarrow
0$ can then be safely taken. Furthermore, to keep the virtues of the
``quantum'' mass-independent scheme, we should use a mass-independent
scheme also for the thin-brane divergences.

This kind of effective theory can in principle describe any sensible
fundamental theory defined in manifolds with branes. These can be
classified into universality classes, with theories within the same
class being described by the same effective theory at lowest
order. Sometimes a small perturbation in the 
ultraviolet of the fundamental theory can turn out to be relevant and
bring the theory to a different universality class. We will show
examples of this situation below.

In this paper we are mainly interested in field theories in more than
four 
dimensions with branes. Extra-dimensional quantum field theories are 
non-renormalizable, so they are necessarily effective theories, even
in the absence of branes. Because of 
the power-law running of the couplings, the cutoff $\Lambda_0$
cannot be much larger than the compactification scale $L^{-1}$ if we
are to stay in a perturbative regime and, at the same time, reproduce
the observed gauge couplings.
On the other hand, the
substructure of the branes is often assumed to be described by some
fundamental theory such as string theory. In this case, they belong
to the class with $\eps_0^{-1} \gtrsim \Lambda_0=\Lambda$.
In extra dimensions,
non-renormalizability implies that an infinite number of localized
operators are generated. At leading order in the low-energy expansion
they consist of localized mass terms, kinetic terms and marginal
interactions~\cite{Georgi:2000ks,Cheng:2002iz}. Some of these brane 
terms, in turn, give rise to singularities and a loss of
perturbativity in the thin-brane limit, unless they are
subtracted. Such a 
renormalization has been proposed and studied
in~\cite{Goldberger:2001tn} for branes of codimension 2 with localized
mass terms, and in~\cite{Lewandowski:2001qp,delAguila:2003bh} for
branes of codimension 1 with 
localized derivative (kinetic) terms. This renormalization is usually
dubbed ``classical'' because it is required already at tree level in
the presence of tree-level brane terms.

Our purpose here is to study the effect of brane and
bulk operators to second order in perturbation theory (up to
$\Lambda^{-2}$), 
and to
give a simple renormalization prescription, which can be used in
phenomenological 
calculations with thin branes. Furthermore, we want to check that the
results obtained with this prescription are physical, in the sense
that they agree with the ones given by a more fundamental theory
incorporating a microscopic description of the branes. We consider
theories with plane parallel branes of codimension~1 and, for
definiteness, restrict to the orbifold $M_4\times S^1/Z_2$, which we
parametrize (in the 
``upstairs'' picture) by $x^\mu$, $\mu=0,1,\ldots,3$ and $y=x^5 \in
(-\pi R, \pi R]$. Our branes are the fixed points of the $Z_2$ action,
  located at $y=0$ and $y=\pi R$. Therefore, they are non-dynamical
  objects and we do not need to include their fluctuations
  (``branons'', see
  \cite{Sundrum:1998ns,Dobado:2000gr,Cembranos:2001rp}) in 
  the effective theory. We study the free theory, which is
  already non-trivial, and concentrate on the
kinetic terms, which are the most 
relevant in phenomenology. 
We give a general
basis of independent operators which, in principle, can describe to a
certain order any ultraviolet completion with the assumed symmetries
in this sector.
Then, we study the impact of these
operators on the KK spectrum in a perturbative
calculation. Finally, we 
match the general (free) effective theory to a specific completion:
deconstructed orbifolds~\cite{decorbs}. The fact that this matching is
possible is a check of the validity of the effective framework, of
classical renormalization, and of our particular prescription.  The
description of the 
free part of brane fields is straightforward in flat space for plane
branes, so we will 
focus mostly on bulk fields\footnote{Brane fields
  contribute to brane-localized free terms for bulk 
  fields via quantum corrections. At the classical level, they can
  contribute as well if there is mass or kinetic mixing with the bulk
  fields. Actually, we will consider below one case in which a brane-bulk
  fermionic mass mixing has dramatic effects. Elsewhere, diagonal mass
  and kinetic matrices are assumed.}. 
We choose to work in the parent theory with fields that have well-defined
orbifold parity, rather than in the interval with boundary
conditions.


\section{Renormalization}

We want an effective theory with branes of vanishing width,
\ie, with the brane-localized terms proportional to
Dirac delta functions. We will work formally with these representations
whenever possible, and only resort to an
intermediate regularization of the delta functions to study certain
finite but ambiguous terms in Section~\ref{oddodd}. Since products of
these delta functions 
appear in perturbation theory, even classically, a subtraction
procedure is in order. 
We propose a simple prescription to perform the subtractions: 
define all products of delta functions or their derivatives
as identically zero. This applies to
products of deltas both in the action and in the calculations of
amplitudes (or KK reduction), and determines a mass-independent
renormalization scheme, which we call {\bf analytical
  renormalization}. The reason for 
the term ``analytical'' is 
that this prescription would follow from 
a regularization by some sort of analytical continuation similar to
dimensional regularization, plus minimal subtraction. In practice,
however, no explicit realization is needed in our calculations for
plane branes of codimension 1, where the thin-brane divergences are
power-like. Nevertheless, as we discuss below, an explicit
  regularization could be useful to deal with finite ambiguous
  contributions, which appear also in the plane case. Moreover,
  logarithmic divergences appear in codimension 
2~\cite{Goldberger:2001tn,Dudas:2005vn,Dudas:2005gi}, and when 
the branes are curved, extra localized terms proportional
to their extrinsic 
curvature arise after careful regularization of the
singularities~\cite{McAvity:1992hy}.
Therefore, a refinement of analytical renormalization
(may be some form of differential
renormalization~\cite{Freedman:1991tk}), or an explicit analytical
regularization, is required in these and other more general
situations.  Our prescription (and the resulting renormalization
scheme) presents several advantages: 
\begin{itemize}
\item It is extremely simple, as no
  intermediate regularization of the 
  Dirac deltas is necessary and no explicit counterterms have to
  be computed. 
\item It does not introduce any dimensionful regulator which could
  interfere with the expansion in inverse powers of $\Lambda$. 
\item As we show below, it allows the elimination of some brane terms
  via field redefinitions.  
\item It preserves supersymmetry, at least in known examples.
\end{itemize}
Observe that even if the thin-brane divergences signal a
dependence of ultraviolet details, we can cancel them completely
since the relevant information is encoded in the finite coefficients
of operators with at most a single delta function.
Putting all products of (derivatives of) deltas to zero is equivalent
to exactly cancelling these products by counterterms in the
renormalized action, as proposed in~\cite{delAguila:2003bh}, but it is
simpler, 
as many terms in amplitudes or in the action can be discarded from the
beginning. 
For instance, and in relation to the last point, it is known
that in supersymmetric theories with branes higher-order terms with
products 
of delta functions have to be included in the action to preserve
supersymmetry~\cite{Mirabelli:1997aj,delAguila:2003bh}.
But if analytical renormalization is 
used in all calculations, these terms can (and should) be
omitted. This is possible because
products of deltas are also renormalized to zero in the on-shell
supersymmetry transformations. Although discarding the divergent
terms is just a devise to save work in the cases in which they
eventually cancel~\cite{Mirabelli:1997aj,Delgado:2001xr}, it does have
physical implications in truly divergent
situations~\cite{delAguila:2003bh}, where it amounts to a
renormalization. In particular, the free action for a supersymmetric
boson with brane terms, which includes an infinite series of higher-order
terms~\cite{delAguila:2003bh}, is equal to the na\"{\i}ve one after
renormalization. This means that the latter, when combined with the
free fermionic action, is supersymmetric with our prescription.
We shall exploit this fact below.

Analytical renormalization allows us to eliminate many operators in
the effective action using field redefinitions. Consider for instance a
kinetic Lagrangian for fermions with general (lowest-order) brane
terms at one of the fixed points:
\begin{align}
\Lag = & (1+a^\L \delta_0) \bchi_\L i \dsl \chi_\L + (1+a^\R 
\delta_0 ) \bchi_\R i \dsl \chi_\R - \half (1 + b^\L \delta_0) (\bchi_\L
\d_5 \chi_\R + (\d_5 \bchi_\R) \chi_\L) \nn
& \mbox{} + \half (1 + b^\R \delta_0) (\bchi_\R
\d_5 \chi_\L + (\d_5 \bchi_\L) \chi_\R) \, ,
\end{align}
with $\delta_0 = \delta(x^5)$ and $\chi_{\L,\R}$ the left-handed and
right-handed chiral projections of five-dimensional (four-component)
Dirac spinors. The brane kinetic terms with derivatives normal to the
branes (``orthogonal'' brane terms), 
with coefficients $b^{\L,\R}$, give rise to thin-brane
singularities in the classical
propagator~\cite{delAguila:2003bh}. Performing a 
field redefinition $\chi_{c} = h_c \psi_c$ with 
$h_c=(1+\frac{b^L+b^R}{2} \delta_0)^{-\frac{b^c}{b^L+b^R}}$ and
  $c=\L,\R$, the free Lagrangian is written as
\be
\Lag =  (1+a^\L \delta_0) h_\L^2 \bpsi_\L i \dsl \psi_\L + (1+a^\R
\delta_0 ) 
h_\R^2 \bpsi_\R i \dsl \psi_\R  - \bpsi_\L
\d_5 \psi_\R + \bpsi_\R \d_5 \psi_\L  \label{redefLag}
\, .
\ee
We have traded the orthogonal brane kinetic terms for parallel
brane kinetic terms (those without normal derivatives) 
times singular expressions. This makes the
divergences associated to orthogonal brane terms apparent. But with
analytical renormalization, \refeq{redefLag} reduces to
\be
\Lag =  (1+\bar{a}^\L \delta_0) \bpsi_\L i \dsl \psi_\L +
(1+\bar{a}^\R 
\delta_0 ) \bpsi_\R i \dsl \psi_\R  - \bpsi_\L
\d_5 \psi_\R + \bpsi_\R \d_5 \psi_\L
\, ,
\ee
with $\bar{a}^c = a^c - b^c$, which contains only
non-singular parallel brane terms. This is equivalent to performing 
the first-order field redefinitions of Ref.~\cite{delAguila:2003bh}
and discarding all the higher-order terms that are generated.  


\section{KK decomposition of renormalized effective theories}

Next we write down general free effective Lagrangians for massless
fermions, scalars and gauge bosons to order  $\Lambda^{-2}$ and
perform the KK reduction to the same order.
We impose 4D Lorentz
invariance in the directions parallel to the branes. We also
impose the full 5D Lorentz invariance in the bulk at zeroth
order\footnote{Note that this is automatic in the free theory if there
is only one particle, as can be seen by a redefinition of the
coordinate $y$.}, but allow for its breaking in higher-order bulk
operators. This is necessary to describe ultraviolet completions
breaking this symmetry, such as deconstruction, and can be
useful in model building~\cite{Panico:2005dh}. In the following we
will refer to the brane terms, \textit{i.e.} to operators with a delta
function, as odd--odd when they involve products of odd functions and
as even--even otherwise. The operators are always invariant, so the
number of odd factors must be even.
Note that odd--odd operators are ambiguous: they vanish formally
but can be non-zero if the delta functions are regularized. As we
shall see in the next subsections, they do not contribute to second order. We
discuss their impact on higher-order corrections in
Section~\ref{oddodd}.

\subsection{Fermions}

In the fermionic case we also
allow for operators proportional to the background
$\sigma(y)=\mbox{sign}(y)$, because they can mimic the effect
of a Wilson term in deconstruction, and for 
a breaking of chiral invariance (without masses). 
The free fermion Lagrangian can be written as $\Lag_f = \Lag_f^{(0)} +
\frac{1}{\Lambda} \Lag_f^{(1)} + \frac{1}{\Lambda^2} \Lag_f^{(2)} +
\ldots$, where
\begin{align}
& \Lag_f^{(0)} = \bpsi (i \dsl - \gamma_5 \d_5) \psi \,  , \\
& \Lag_f^{(1)} = \kappa_1 \sigma (\d_5 \bpsi) \d_5 \psi + a_I^\R \delta_I
  \bpsi_\R i \dsl \psi_\R + a_I^\L \delta_I
  \bpsi_\L i \dsl \psi_\L \, , \\
& \Lag_f^{(2)} = \kappa_2 \bpsi_\L \d_5^3 \psi_\R + \xi_I \sigma \delta_I
  (\d_5 \bpsi_\L)\d_5 \psi_\R + \eta_I^\L \sigma \delta_I \bpsi_\L \d_5^2
  \psi_\R +  \eta_I^\R \sigma \delta_I \bpsi_\R \d_5^2
  \psi_\L  + \hc \, .  \label{fermlag}
\end{align}
Here, $\dsl = \d_\mu \gamma^\mu$, $\delta_I=\delta(y-R
I)$, with $I=0,\pi$ labelling the positions 
of the fixed points, and sums over repeated indices $I$ are
understood. We choose $\psi_\R$ ($\psi_\L$) to be even (odd) under the
orbifold parity. All the parameters, except $\Lambda$, are
dimensionless. Several possible operators have  
been eliminated by integration by parts, use of the zeroth-order
equations of motion (or equivalently, perturbative field
redefinitions) and analytical renormalization. 
The values $\kappa_1=\kappa_2=0$ correspond to 5D Lorentz
invariance 
in the bulk. 
We have chosen a basis
of operators which leads to a convenient KK reduction, such that the
resulting 4D theory has no higher-derivatives in the
kinetic term. Indeed, if we
expand $\psi_{\L,\R}(x,y) = \sum_n f^{\L,\R}_n(y) \Psi_{\L,\R \;
  n}(x)$ and 
take $f^{\L,\R}_n$, $m_n$ to be the eigenvectors and eigenvalues of the
generalized eigenvalue problem
\begin{align}
& \big[-\d_5 + \frac{\kappa_1}{\Lambda} \d_5 \sigma \d_5 +
  \frac{\kappa_2}{\Lambda^2} \d_5^3 \big] f^\L_n
  = m_n (1+\frac{a_I^\R}{\Lambda} \delta_I) f_n^\R,  \nn
& \big[\d_5+\frac{\kappa_1}{\Lambda} \d_5 \sigma \d_5 -
  \frac{\kappa_2}{\Lambda^2} \d_5^3 \big] f^\R_n = m_n
   f_n^\L \, ,  \label{eigensystem}
\end{align} 
with normalization 
\be
1=\int_{-\pi R}^{\pi R} \mathrm{d} y \,
\left(1+\frac{a_I^{\R}}{\Lambda} \delta_I \right) 
(f_n^{\R})^2 =  \int_{-\pi R}^{\pi R} \mathrm{d} y \, 
(f_n^{\L})^2
\, , 
\label{normalization}
\ee
the free Lagrangian reduces to
\be
\Lag_f = \sum_n  \bar{\Psi}_n (i \dsl - m_n) \Psi_n \, ,
\ee
with $\Psi_n=\Psi_{\L\,n}+\Psi_{\R\,n}$, and $\Psi_{\L\,\bar{n}}=0$
($\Psi_{\R\,\bar{n}}=0$)   
for possible right-handed (left-handed) modes with $m_{\bar{n}}=0$. In
writing the   
eigensystem~\refeq{eigensystem} we have used the fact that the terms
in the action with coefficients $a_I^\L$, $\xi_I$, $\eta_I^\L$ and
$\eta_I^\R$ do 
not contribute to second order, as they vanish when the (continuous)
zeroth-order wave functions are used. Then, to second order we can
safely work with strict delta functions and the calculation is
straightforward: at each order we solve the bulk equation and apply
the boundary (``jump'') conditions found by integrating around the
fixed points. 

Expanding the KK masses and wave functions in
$1/\Lambda$ we find a flat right-handed zero mode plus a tower with
KK masses
\be
m_n = \frac{n}{R} \left[1 + A \frac{1}{R \Lambda} + (A^2 + B n^2)
  \frac{1}{(R \Lambda)^2} \right] + \ldots  \, ,
  \label{fermionmass} 
\ee
where $A=-\frac{a^R_0+a^R_\pi}{2\pi}$,
$B=\frac{\kappa_1^2}{2}+\kappa_2$ and $n=1,2,\ldots$ The structure
of~\refeq{fermionmass}, with a piece proportional to $n$ depending on a
single number $A$ and another piece proportional to $n^3$, which
appears at second order, is a consequence of the symmetries we have
imposed on the effective Lagrangian, of the fact that the expansion in
local operators is controlled by a single scale $\Lambda$, and of
analytical renormalization, which does not mix up this ordering. The
wave functions of massive modes in the 
fundamental region have the structure
\be
f^c(y) = P_1^c(y) \cos\left(\frac{n}{R} y\right) + P_2^c(y)
\sin\left(\frac{n}{R} y\right),  
\label{wffermion} 
\ee
with $P^c_{1,2}$ polynomials of second degree (to second order in
$\Lambda^{-1}$). Here we write them to first order only:
\begin{align}
& P_1^\R(y) = N_0 + N_1 \frac{1}{R \Lambda} \, , \nn 
& P_2^\R(y) = - N_0 \frac{n}{R \Lambda} \left(\frac{a_0^\R}{2} + A
    \frac{y}{R} \right) \, , \nn 
& P_1^\L(y) =- N_0 \frac{n}{R \Lambda} \left(\frac{a_0^\R}{2} +
    \kappa_1 + A   \frac{y}{R} \right) \, , \nn
& P_2^\L(y) = -N_0 - N_1 \frac{1}{R \Lambda} \, ,
    \label{wffermion1storder} 
\end{align}
with $N_0$ and $N_1$ perturbative normalization constants.
Observe that the parameter $\kappa_1$
only appears in the KK masses in the combination
$\frac{\kappa_1^2}{2}+\kappa_2$ to 
second order. On
the other hand, the first-order wave function for the left-handed 
component depends on $\kappa_1$ but not on $\kappa_2$; therefore, the
operator with coefficient $\kappa_1$ is not redundant. The wave
functions also distinguish $a_0^\R$ from $a_\pi^\R$ at first order.

\subsection{Scalars}

For a massless complex scalar, after integration by parts, field
redefinitions and analytical renormalization, the effective Lagrangian
to second order reads $\Lag_s = \Lag_s^{(0)} +
\frac{1}{\Lambda} \Lag_s^{(1)} + \frac{1}{\Lambda^2} \Lag_s^{(2)} +
\ldots$, with
\begin{align}
& \Lag_s^{(0)} = \phi^\dagger (-\Box + \d_5^2) \phi \,  , \\ 
& \Lag_s^{(1)} = a_I \delta_I \phi^\dagger \Box \phi - c_I \delta_I
  (\d_5 \phi^\dagger) \d_5 \phi \, , \\
& \Lag_s^{(2)} = \kappa \phi^\dagger \d_5^4 \phi \, ,
  \label{scalarlag}
\end{align}
and $\Box=\d_\mu \d^\mu$. We have not included terms proportional to
$\sigma$ in the scalar case, as they are not required to reproduce the
results in deconstruction. A possible orthogonal brane kinetic term
$b_I \delta_I \phi^\dagger \d_5^2 \phi + \hc$ has been absorbed into
the $a$-term, using field redefinitions and analytical renormalization. If 5D
Lorentz invariance is preserved in the bulk, then $\kappa=0$. 
The KK reduction is performed by expanding $\phi(x,y)=\sum_n f_n(y)
\Phi_n(x)$ with $f_n$ the eigenfunctions of the eigenvalue problem
\be
\O_s f_n =
-m_n^2 (1+ a_I \delta_I) f_n  \, , \label{scalareigen}
\ee 
normalized as
\be
1=\int_{-\pi R}^{\pi R} \mathrm{d} y \, (1+a_I \delta_I) f_n^2 \, . 
\ee
The operator in \refeq{scalareigen} is $\O_s=\d_5^2 +
\frac{c_I}{\Lambda} \d_5 \delta_I \d_5 + \frac{\kappa}{\Lambda^2}
\d_5^4$. 
Using analytical renormalization it is possible to reduce this problem
to a fermionic one. Indeed, after renormalization, the
``supersymmetric'' operator 
$\tilde{\O}_s= - \O_f^\dagger \O_f$, with  $\O_f=-(1+ \frac{c_I }{2
\Lambda} \delta_I) \d_5 - \frac{\kappa}{2\Lambda^2} \d_5^3$, is
identical to $\O_s$, to second order. On the
other hand, if $f_{1\,n}$, $f_{2\,n}$ and $m_n$ are solutions of the
fermionic eigensystem 
\begin{align}
& \O^\dagger_f f_{2\,n} = m_n (1+a_I \delta_I) f_{1\, n} \, , \nn
& \O_f f_{1\, n} = m_n f_{2\, n} \, ,  \label{fermion2}
\end{align}
then $f_n=f_{1n}$ and $m_n$ are obviously solutions of $\tilde{\O}_s
f_n = -m_n^2 (1+a_I \delta_I) f_n$, and hence of~\refeq{scalareigen}.
Finally, by field redefinitions and analytical renormalization,
we can eliminate the derivatives of delta functions and reduce
\refeq{fermion2} to \refeq{eigensystem} with $a^\L_I=a_I$, $a^\R_I=-c_I$,
$\kappa_1=0$ and $\kappa_2=-\kappa/2$. The role of $f_{1n}$ is played
by $f^\L_n$. Therefore, the results obtained above give the
KK decomposition of an odd scalar. In particular, there is no zero
mode. For an even scalar we can
change $\O_f \rightarrow \O_f^\dagger$ above, which leads to $f_{1
  n}=f^\R_n$ and parameters $a^\L_I=-c_I$, $a^\R_I=a_I$, $\kappa_1=0$ and
$\kappa_2=-\kappa/2$. 

We find that the KK masses are given
by~\refeq{fermionmass} with 
$A=-\frac{a_0+a_\pi}{2\pi}$ for an even scalar,
$A=\frac{c_0+c_\pi}{2\pi}$ for an odd one, and $B=-\frac{\kappa}{2}$ in
both cases. Note that the terms with coefficients $c_I$
give a non-trivial contribution for odd scalars, despite
the fact that they do not contribute when treated
non-perturbatively without
renormalization~\cite{delAguila:2003bh,delAguila:2004xd}. 
The wave functions also follow directly from the fermionic ones
$f^{L,R}_n$ in~\refeq{wffermion} and~\refeq{wffermion1storder}, using
the same particular values for the fermionic parameters.

\subsection{Gauge bosons}

The case of gauge bosons is a special case of the scalar one when the
gauge $A_5=0$ is chosen. The main difference is that the most singular
brane terms are forbidden by gauge invariance, but after analytical
renormalization the free Lagrangians are equivalent. Indeed, after
some field redefinitions the most general gauge kinetic Lagrangian to
second order is $\Lag_g = \Lag_g^{(0)} +
\frac{1}{\Lambda} \Lag_g^{(1)} + \frac{1}{\Lambda^2} \Lag_g^{(2)} +
\ldots$, with
\begin{align}
& \Lag_g^{(0)} = -\frac{1}{4} F_{\mu\nu} F^{\mu\nu} - \frac{1}{2}
   F_{\mu 5} F^{\mu 5} \,  , \\ 
& \Lag_g^{(1)} = -\frac{1}{4} a_I \delta_I F_{\mu\nu} F^{\mu\nu} -
   \frac{1}{2} c_I \delta_I F_{\mu 5} F^{\mu 5} 
   \, , \\
& \Lag_g^{(2)} = -\frac{1}{2} \kappa F_{\mu 5} \d_5^2 F^{\mu 5} \, , 
  \label{gaugelag}
\end{align}
which, upon writing $F_{MN}=\d_{[M} A_{N]}$ and putting $A_5=0$, is
identical to the real-scalar version of~\refeq{scalarlag}, with
$\phi\rightarrow A_\mu$. Therefore, the KK reduction of $A_\mu$ gives
the same expressions. 


\section{Matching with fundamental theories}

These results in the effective theory should agree with those
obtained from a more fundamental theory in which the physics around
the fixed points is non-singular. This is the case of weakly coupled
string theory, where the extended nature of the strings softens
the orbifold singularities. In perturbative string theory on an
orbifold, 
some of the contributions of string loop corrections to a given
correlation function are localized around the fixed points. These
contributions are suppressed by powers of the string coupling, and the
localization profile is controlled by the string length $l_s$, which
is the only dimensionful parameter at hand (we are assuming a large
compactification radius, so that finite-size effects are
small). Therefore, even if we start with unresolved orbifolds with
$\eps_0=0$, effectively this is smeared to $\eps_0 \simeq l_s$. These
calculations have been performed 
explicitly for localized gauge-field tadpoles in heterotic string
theory in~\cite{GrootNibbelink:2003zj} (see
also~\cite{Atick:1987gy}--\cite{Antoniadis:2002tr}).
The results are 
in agreement with our dimensional analysis, but a dimensionless factor
arising from normal-ordering constants in
the world-sheet field theory turns out to be crucial as well. We refer
to~\cite{GrootNibbelink:2003zm} for details about the analogies and
differences in 
the string and the field-theory calculations. Other 
``fundamental theories'' in which the physics around the fixed points
is smooth are the field-theoretical orbifold resolutions
of~\cite{Serone:2004yn}, which should be regarded as effective
theories valid 
up to a cutoff $\Lambda_0$ larger than the inverse size of the
resolved fixed points, $\eps_0^{-1}$. In this case, there can be
localized terms involving the curvature and the flux
backgrounds.
 
To be more quantitative, we compare the renormalized effective theory
with another ultraviolet completion: a deconstructed
version~\cite{Arkani-Hamed:2001ca,Hill:2000mu} of the 
orbifold 
$S^1/Z_2$. Deconstructed orbifolds were introduced
in~\cite{Hill:2000mu} (in the fundamental region) and will be studied
in greater detail in~\cite{decorbs} (starting from the parent theory
space). They are renormalizable 
4D theories in which the gauge group is formed by a
product of identical simple groups, with  ``link'' scalar
fields charged under ``neighbouring'' pairs of
group factors. Below the scale  
of the vacuum expectation value of the link scalars $v$, they
are equivalent to 5D field theories whose fifth
dimension is a latticized segment. The lattice spacing is $s=(g
v)^{-1}$, with $g$ a dimensionless coupling, and the radius $R$ of the
discrete $S^1$ is given 
by $\pi R= N s$, where $N$ is the number of
sites in the segment (the fundamental region). Scalar and fermion
fields can also be added at each site to represent bulk scalars and
fermions. We use Wilson fermions to avoid doubling, and fine-tune the
parameters in such a way that chiral invariance is recovered in the
continuum limit. The sites near the
boundary of the interval behave differently from those deep inside
the bulk, and localized terms are generated by quantum
corrections, with coefficients independent of $s$ and $N$. In this
scenario, the fixed points (boundaries) are described by Kronecker,
rather than Dirac, deltas. The effect of the brane tems is
thus regulated by the lattice spacing, which acts as a cutoff in
position space. Therefore, we effectively have
$\eps_0=\Lambda_0^{-1}\equiv s$. In deconstruction, KK reduction
amounts to a diagonalization of the mass matrix arising from the
discrete kinetic term. 
The deconstructed orbifold theory can be
described at energies below $\Lambda=1/s$ by a (classically)
renormalized effective theory in a 
continuous orbifold with cutoff $\Lambda$. This means, in particular,
that the KK masses and the discrete wave functions agree with the
general ones we have found here. To see this, we must expand them to
second order in a Taylor series about $s=0$, keeping $R$ 
fixed. Let us summarize
the results.

For massless fermions, gauge bosons, massless odd scalars
and generic massless even scalars, we find the following KK masses to
second order~\cite{decorbs}: 
\be
m_n = \frac{n}{R} \left[1 + \mathcal{A} \frac{s}{R} +
  \left(\mathcal{A}^2 -  \frac{n^2 \pi^2}{24} \right)
  \left(\frac{s}{R}\right)^2 \right] + \ldots  \, ,
  \label{decomass} 
\ee
where the value of $\mathcal{A}$ depends on the kind of field and is a
function of the coefficients of different operators near
the fixed points in the deconstructed theory. For even gauge bosons
and a fine-tuned class of even scalars---in which certain
combinations of brane coefficients are put to zero---there is a flat
zero mode. On the other hand, for generic even scalars the zero mode
disappears and we find instead two tachyons, one localized at each
brane. Their mass is proportional to the inverse spacing. For
fermions there is a flat chiral zero mode; in some deconstructed
orbifolds, there is in addition one zero mode localized at one of the
branes, which has the same chirality as the bulk mode for
``chiral'' deconstructed orbifolds, and opposite chirality for
``non-chiral'' orbifolds. To next-to-leading order, the wave functions
for massive KK modes of ($Z_2$-even) right-handed massless 
fermions, even gauge bosons and fine-tuned even massless scalars read
\be
f^{(1)}_n = \left(N_0^\prime + N_1^\prime  
  \frac{s}{R}\right) \cos\left(\frac{n y}{R}\right) - N_0^\prime
\frac{n}{R \Lambda}   
  \left(\mathcal{C}_1 + \mathcal{A} \frac{y}{R} \right)
  \sin\left(\frac{n y}{R}\right) \, ,  
\ee
where $y=i s$ and $i$ labels the
sites. On the other hand, the wave functions of the corresponding
(odd) left-handed fermions and 
of odd gauge bosons, odd massless scalars and generic even massless
scalars are
\be
f^{(2)}_n =- N_0^\prime \frac{n}{R \Lambda}
  \left(\mathcal{C}_2 + \mathcal{A} \frac{y}{R} \right) \cos\left(\frac{n
    y}{R}\right) -  \left(N_0^\prime + N_1^\prime \frac{s}{R}\right)
  \sin\left(\frac{n y}{R}\right) \, . 
\ee
Here, $N_{0,1}^\prime$ are normalization constants and $\mathcal{A}$
is the same expression (for each case) as appears in the masses.

All these results can be reproduced by the renormalized effective
theories in the continuum. Indeed, Eq.~\refeq{decomass} neatly matches
the generic expression \refeq{fermionmass} and the first-order
deconstructed wave
functions above have the same form as the ones in~\refeq{wffermion}
and~\refeq{wffermion1storder}. Adjusting the parameters of the
effective Lagrangians, we find exact agreement for KK masses and the
corresponding wave functions in each case. The exact values in terms
of the parameters 
of the different deconstructed models will be given
in~\cite{decorbs}.
Finally, the localized zero modes and tachyons can be
described directly in the effective theory by massless and tachyonic
fields, respectively, living on the branes.
 
Interestingly enough, it turns out that the deconstructed generic even 
scalars are described by odd scalars (plus the brane tachyons) in the
effective theory. This is 
due to the presence of discrete brane operators, which look like
irrelevant from 
na\"{\i}ve power counting, but turn out to be relevant and change
drastically the continuum limit. 
Hence, these theories
belong to the same universality class as that of deconstructed odd
scalars, except for the brane instability. When these operators are
put to zero, the theory stays in the universality class one would have
na\"{\i}vely guessed.
An alternative
  effective description is to use 
  even scalars and add tachyonic boundary masses of the order of
  $\Lambda$. These are Dirac delta well potentials, which localize one  
  mode---the tachyon---at each brane. The
  remaining KK modes are expelled from the branes by
  orthogonality\footnote{A big non-tachyonic boundary mass also gives
    rise to Dirichlet boundary conditions but does not localize any
    mode.}. Even though changing the parity of the field is
  simpler, the description of this effect by explicit relevant
  operators has several advantages. First, their behaviour under
  important symmetries can be studied. Second, their coefficients
  are dimensionful, and thus naturally of the order of the
  cutoff. This shows that effective Dirichlet boundary conditions for
  even scalars are
  natural. Third, the coefficients can be fine-tuned to be much
  smaller than the cutoff, in order to reproduce the fine-tuned
  scenario with effective Neumann boundary conditions. And fourth,
  their running in the effective theory can be studied with standard
  methods. 

It might be thinked that the non-decoupling effects should be related
to the instabilities. However, there are stable examples in which this
change of boundary conditions in the infrared is observed. For
instance, in the non-chiral class of deconstructed
fermions, if we do not fine-tune the mass and Wilson
term near the boundaries (as has been assumed so far), both zero modes 
combine to form a massive Dirac mode~\cite{decorbs}. 
The full set of KK masses is
then given by 
\be
m_n = \frac{n+1/2}{R} \left[1 + \mathcal{B} \frac{s}{R} +
  \left(\mathcal{B}^2 - 
  \frac{(n+1/2)^2 \pi^2}{24} \right)
  \left(\frac{s}{R}\right)^2 \right] + \ldots  \, ,
  \label{decomass2} 
\ee
while the wave functions of the right-handed fermion are
\begin{align}
f^{(3)}_n = & \left(N_0^\prime + N_1^\prime 
  \frac{s}{R}\right) \cos\left(\frac{(n+1/2) y}{R}\right) \nonumber \\
  & \mbox{}- N_0^\prime
  \frac{n+1/2}{R \Lambda} 
  \left(\mathcal{C}_3 + \mathcal{B} \frac{y}{R} \right)
  \sin\left(\frac{(n+1/2) y}{R}\right) \, ,  
\end{align}
to first order. The same form of the masses and wave functions, plus
one localized tachyon, is obtained for
deconstructed even scalars with the mentioned operators adjusted to
zero only near the boundary $y=0$. This
behaviour can be precisely matched to an effective theory in which
different boundary conditions are used at the two boundaries for each
chiral component (Neumann--Dirichlet at lowest order). In the
orbifold formalism in the ``parent'' space, this can be achieved by
allowing for a twist such that the field has antiperiodic boundary
conditions, or equivalently, using an orbifold $S^1/(Z_2\times
Z_2^\prime)$. As in the scalar case, an alternative description
preserving the original parity of the fields is possible, if relevant
operators are included. Specifically, the operator doing 
the job is a mass mixing between the right-handed bulk fermion and a
left-handed localized mode at one of the branes (which was present in 
the purely massless case). Its coefficient has mass dimension 1/2, and
is naturally of order $\Lambda^{1/2}$. This operator, which breaks the
chiral invariance of the bulk fermion, mimics faithfully
the physical mechanism involved in the change of the continuum
boundary conditions of the deconstructed theory. More details will be
given in~\cite{decorbs}.


\section{Odd--odd operators}
\label{oddodd}

We have seen that operators containing a delta function
times a product of odd functions (odd--odd brane terms) do not have
any effect to second order in the effective theory. The same holds in
deconstruction, but these terms
start contributing at third order (in $s$).
If we want to reproduce this effect with odd--odd
terms in the effective theory we need to regularize the delta
functions, for otherwise these terms are not well defined (they vanish
formally). This typically involves a dimensionful
regulator, which could mix with the expansion in $\Lambda$ and
reintroduce the thin-brane singularities. We have found explicitly that the
``point-splitting'' regulator introduced in~\cite{Csaki:2003sh} does
not have 
this problem. This regularization consists in shifting the support of
the delta functions a distance $\eps$ away from the fixed
points. Of course, it would be simpler if we could find a prescription
to deal with these terms without introducing any dimensionful
regulator. This could follow from an explicit analytical
regularization and is under investigation. At any rate, using
point-splitting regularization and taking $\eps\rightarrow 0$ at the
end, we find 
that the third-order contribution to the 
fermion KK masses of the (odd--odd) terms proportional to
$a^\L_{0,\pi}$ is 
\be
-\frac{n^3}{32 \pi R^4 \Lambda^3}\, \big[(a^\L_0)^2 a^\R_0 +
  (a^\L_\pi)^2 a^\R_\pi\big] \, . 
\label{thirdorder}
\ee
The numerical coefficient is regularization dependent. This is not a
problem, as this dependence can be absorbed 
into the renormalized couplings $a^\L_I$.
Observe that the contribution of $a^\L_I$ in~\refeq{thirdorder}
vanishes if $a^\R_I$ (with the 
same $I$) does. The reason for this is that the odd--odd terms
contribute only when the fields are discontinuous at the branes, and
this discontinuity is induced at lower orders by $a^\R_I$. 
The dependence on $n$ and $R$ matches the one of 
the corresponding contribution obtained in deconstruction. However, in
this case the even--even brane terms, which give rise to $a^\R_I$ in the
continuum, do not need to be turned on to have a non-vanishing
contribution of the third-order odd--odd term. The effective theory can
also reproduce such correction with $a^\R_I=0$ by means of
the third-order operator $\delta_I \bpsi_\R 
\d_y^3 \bpsi_\L$, which gives a contribution with the same $R$ and $n$
dependence as~\refeq{thirdorder}. This shows that, at least as far as
the KK masses are concerned, the operators with coefficient $a^\L_I$
are redundant to third order. 

One might speculate that this is a general property of the free
effective theory, \ie that the
effect of any odd--odd term, to all orders, can be absorbed into
higher-order even--even terms. We have not found a field
redefinition 
showing this, and in fact field redefinitions preserving the orbifold
parity will not mix odd--odd with even--even operators. However the
possibility that odd--odd terms be redundant agrees with the idea that
the free brane 
operators simply determine the boundary conditions outside the core of
the brane~\cite{Lewandowski:2001qp}, and an arbitrary boundary
condition can be 
imposed by adjusting the value of the even--even operators. From this
argument, however, it does not follow that the correct dependence 
on the KK number at each order will be reproduced.
On the other hand, in the interacting effective theory the odd--odd
operators run in general with the renormalization group scale,
and this running would have to be 
incorporated into the even--even operators. This would require an
explicit relation between even--even and odd--odd operators. 
These issues deserve further study.


\section{Conclusions}

We have argued, and showed explicitly in particular
models, that effective theories in extra-dimensional spaces with
infinitely thin defects are a good description of more fundamental
theories in which the defects can have some structure. A
renormalization 
procedure is necessary to take care of the divergences which appear in
the thin-brane limit and we have proposed a simple renormalization
prescription for plane branes of codimension 1, analytical
renormalization, which defines these
divergences as vanishing. We have shown that, 
in this scheme, even--even orthogonal brane kinetic terms can
be completely eliminated by a field redefinition. Odd--odd parallel
and orthogonal terms can also be 
disregarded to second order and maybe higher.
As a matter of fact, only even--even parallel brane kinetic
terms (besides mass brane terms) are customarily taken into account in
phenomenological 
fits~\cite{Carena:2002me}--\cite{Carena:2004zn} and
model building~\cite{Scrucca:2003ra}--\cite{Panico:2005ft}. Our
results imply that, in 
the framework of a classically renormalized effective theory, this
is consistent and does not entail a loss of generality. Moreover, it
agrees with completions such as deconstructed orbifolds.
The less common works including orthogonal terms use fat
branes~\cite{Kolanovic:2003am} or treat the orthogonal brane 
terms perturbatively to first order~\cite{Cheng:2002iz}, which is
non-singular. The first possibility reduces to the renormalized
effective theory at scales lower than the physical width of the
branes. Regarding the second one, we have shown
that the first-order results in~\cite{Cheng:2002iz} are not spoiled by
divergent higher-order contributions if perturbative renormalization
is implemented. 

It should be observed that all our results are 
perturbative in the derivative expansion and assume that dimensionless
couplings are of order 1. However, parallel brane kinetic
terms with coefficients larger than $R$ are often invoked. This puts
the theory in a non-perturbative regime. Even though the effect of the
large parallel kinetic terms (of first order, formally) can be
resummed to all orders, the result may 
be changed by contributions of higher-order operators,
which can be of the same size in principle.
It can still be assumed that all
higher-order terms vanish or have small coefficients, although this choice
is not protected by any symmetry. From this point of view, the
calculations with small brane terms, as those in universal extra
dimensions~\cite{Cheng:2002iz,Cheng:2002ab}, are more robust.

On the other hand, we have seen in explicit examples that the
renormalized effective 
formalism can also describe, perturbatively, scenarios in which the
infrared behaviour 
is abruptly changed when certain operators in the
fundamental theory are turned on. This effect is reproduced in the
effective theory  
by relevant operators, with coefficients which are naturally of the
size of the cutoff---but can be smaller if the symmetry is
enhanced when they vanish. We remark that similar effects would be
produced by certain irrelevant operators in the effective theory
at the regularized level, if the divergences were not substracted. From
this point of view, classical renormalization in 
a mass-independent scheme can be understood as a reorganization of the
effective theory, such that the impact of each operator is controlled
by the size of its coefficient. This allows us to work at a fixed
order consistently, since the contribution of higher-order operators
to a given observable is guaranteed to be smaller (as long as their
dimensionless coefficients are of order 1).
On the other hand, the relevant operators can be alternatively
represented by different orbifold field transformations (or boundary
conditions), at least at the classical level.

In this paper we have focussed on the free local sector of
effective theories and their deconstructed completions, 
although we have in mind the possibility that some of the operators
may be partially or fully induced by quantum corrections when
interactions are included. Before concluding, let us make a few
comments on the interactive theory.
As long as a complete set operators is included in
the effective theory, we should be able to reproduce the interactions
of completions such as the deconstructed models. The tree graphs will
have in general new thin-brane singularities, which can be subtracted
with analytical renormalization. odd--odd terms
appear too, and can be treated as discussed above\footnote{Bulk
higher-derivative operators giving rise to singularities and
ill-defined products can
be redefined away using the classical equations of motion, as shown
in~\cite{Lewandowski:2001qp}.}.
In~\cite{Symanzik:1981wd,McAvity:1992fq} complete quantum computations
in renormalizable field theories of dimension 4 with boundaries have
been performed, and the corresponding 
renormalization group equations have been studied. The techniques
in these references can be applied to theories of dimension higher
than 4, although in this case the singularities will be more severe and
more counterterms will be required because of
non-renormalizability. There are many examples of loop calculations in
extra dimensions with branes in the literature.
See for
instance~\cite{Georgi:2000ks,Cheng:2002iz,Delgado:2001xr,Cheng:2002ab,DaRold:2003yi,
  Contino:2001nj,Barvinsky:2005ms,Ponton:2005kx}.
At the quantum level, one must take care of the usual ``quantum'' UV
divergences at coincident points, which 
can be divided into bulk and brane-localized divergences, and also of
the thin-brane ``classical'' divergences we 
have been discussing so far. In our
perturbative effective framework, the quantum divergences should
also be treated in a mass-independent renormalization scheme, such as 
dimensional regularization with minimal subtraction, so that
the renormalization scale $\mu$ only appears inside logarithms, and
does not interfere with the counting of powers of $\Lambda$. In this
case, the localized quantum divergences are proportional to exact
delta functions and can be cancelled by brane counter\-terms in the
effective theory with vanishing brane
width~\cite{Symanzik:1981wd,McAvity:1992fq,Georgi:2000ks} (for
examples showing the smooth profile of divergences 
with a hard 
cutoff, see~\cite{DaRold:2003yi,Perez-Victoria:2004ef}). The
thin-brane divergences, on the other hand, which appear typically in
one-particle reducible (sub)diagrams, 
can be subtracted using our prescription. In
general, the perturbative renormalization of both classical and quantum
divergences in extra-dimensional theories with branes can be
organized along the lines of Appendix~B in~\cite{Symanzik:1981wd}.
Finally, if a matching with a fundamental theory is performed, this
should be carried out, as usual, at a renormalization scale
$\mu=\Lambda$. Then, the renormalization group equations of the
effective theory can be used to calculate processes at lower
energies.

\subsection*{Acknowledgements}
It is a pleasure to thank M.~L\"uscher for enlightening discussions.
We would also like to thank C.~P.~Burgess, M.~Carena, J.~R.~Pel\'aez, L.~Pilo,
E.~Pont\'on, R.~Rattazzi, M.~Serone, T.~Tait, C.~Wagner and A.~Wulzer for
useful conversations. This work has been partially supported by 
URA under Contract No. DE-AC02-76CH03000 with the US DOE, by MEC
(FPA 2003-09298-C02-01), and by Junta de Andaluc\'{\i}a (FQM 101).


\begin{thebibliography}{99}

\bibitem{decorbs} 
F.\ del Aguila, M.\ P\'erez-Victoria and J.\ Santiago, in
preparation. 

\bibitem{Weinberg:1978kz}
  S.~Weinberg,
  Physica A {\bf 96} (1979) 327.

\bibitem{Leutwyler:1993iq}
  H.~Leutwyler,
  Ann. Phys.\  {\bf 235} (1994) 165
  [arXiv:hep-ph/9311274].

\bibitem{Aghababaie:2003iw}
  Y.~Aghababaie and C.~P.~Burgess,
  Phys.\ Rev.\ D {\bf 70} (2004) 085003
  [arXiv:hep-th/0304066].

\bibitem{Graham:2003ib}
  N.~Graham, R.~L.~Jaffe, V.~Khemani, M.~Quandt, O.~Schroeder and H.~Weigel,
  Nucl.\ Phys.\ B {\bf 677} (2004) 379
  [arXiv:hep-th/0309130].

\bibitem{Symanzik:1981wd}
  K.~Symanzik,
  Nucl.\ Phys.\ B {\bf 190} (1981) 1.

\bibitem{McAvity:1992fq}
  D.~M.~McAvity and H.~Osborn,
  Nucl.\ Phys.\ B {\bf 394} (1993) 728
  [arXiv:cond-mat/9206009].

\bibitem{Georgi:2000ks}
  H.~Georgi, A.~K.~Grant and G.~Hailu,
  Phys.\ Lett.\ B {\bf 506}, 207 (2001)
  [arXiv:hep-ph/0012379].

\bibitem{Cheng:2002iz}
  H.~C.~Cheng, K.~T.~Matchev and M.~Schmaltz,
  Phys.\ Rev.\ D {\bf 66} (2002) 036005
  [arXiv:hep-ph/0204342].

\bibitem{Goldberger:2001tn}
  W.~D.~Goldberger and M.~B.~Wise,
  Phys.\ Rev.\ D {\bf 65} (2002) 025011
  [arXiv:hep-th/0104170].

\bibitem{Lewandowski:2001qp}
  A.~Lewandowski and R.~Sundrum,
  Phys.\ Rev.\ D {\bf 65} (2002) 044003
  [arXiv:hep-th/0108025].

\bibitem{delAguila:2003bh}
  F.~del Aguila, M.~P\'erez-Victoria and J.~Santiago,
  JHEP {\bf 0302} (2003) 051
  [arXiv:hep-th/0302023].

\bibitem{Sundrum:1998ns}
  R.~Sundrum,
  Phys.\ Rev.\ D {\bf 59} (1999) 085010
  [arXiv:hep-ph/9807348].

\bibitem{Dobado:2000gr}
  A.~Dobado and A.~L.~Maroto,
  Nucl.\ Phys.\ B {\bf 592} (2001) 203
  [arXiv:hep-ph/0007100].

\bibitem{Cembranos:2001rp}
  J.~A.~R.~Cembranos, A.~Dobado and A.~L.~Maroto,
  Phys.\ Rev.\ D {\bf 65}, 026005 (2002)
  [arXiv:hep-ph/0106322].

\bibitem{Dudas:2005vn}
  E.~Dudas, C.~Grojean and S.~K.~Vempati,
  arXiv:hep-ph/0511001.

\bibitem{Dudas:2005gi}
  E.~Dudas, C.~Papineau and V.~A.~Rubakov,
  arXiv:hep-th/0512276.

\bibitem{McAvity:1992hy}
  D.~M.~McAvity and H.~Osborn,
  J.\ Phys.\ A {\bf 25} (1992) 3287.


\bibitem{Freedman:1991tk}
  D.~Z.~Freedman, K.~Johnson and J.~I.~Latorre,
  Nucl.\ Phys.\ B {\bf 371} (1992) 353.

\bibitem{Mirabelli:1997aj}
  E.~A.~Mirabelli and M.~E.~Peskin,
  Phys.\ Rev.\ D {\bf 58} (1998) 065002
  [arXiv:hep-th/9712214].

\bibitem{Delgado:2001xr}
  A.~Delgado, G.~von Gersdorff and M.~Quiros,
  Nucl.\ Phys.\ B {\bf 613} (2001) 49
  [arXiv:hep-ph/0107233].

\bibitem{Panico:2005dh}
  G.~Panico, M.~Serone and A.~Wulzer,
  arXiv:hep-ph/0510373.

\bibitem{delAguila:2004xd}
  F.~del Aguila, M.~P\'erez-Victoria and J.~Santiago,
  Nucl.\ Phys.\ Proc.\ Suppl.\  {\bf 135} (2004) 295
  [arXiv:hep-ph/0410082].

\bibitem{GrootNibbelink:2003zj}
  S.~Groot Nibbelink and M.~Laidlaw,
  JHEP {\bf 0401} (2004) 004
  [arXiv:hep-th/0311013].

\bibitem{Atick:1987gy}
  J.~J.~Atick, L.~J.~Dixon and A.~Sen,
  Nucl.\ Phys.\ B {\bf 292} (1987) 109.

\bibitem{Dine:1987gj}
  M.~Dine, I.~Ichinose and N.~Seiberg,
  Nucl.\ Phys.\ B {\bf 293} (1987) 253.

\bibitem{Poppitz:1998dj}
  E.~Poppitz,
  Nucl.\ Phys.\ B {\bf 542} (1999) 31
  [arXiv:hep-th/9810010].

\bibitem{Antoniadis:2002tr}
  I.~Antoniadis, R.~Minasian and P.~Vanhove,
  Nucl.\ Phys.\ B {\bf 648} (2003) 69
  [arXiv:hep-th/0209030].

\bibitem{GrootNibbelink:2003zm}
  S.~Groot Nibbelink and M.~Laidlaw,
  JHEP {\bf 0401} (2004) 036
  [arXiv:hep-th/0311015].

\bibitem{Serone:2004yn}
  M.~Serone and A.~Wulzer,
  Class.\ Quant.\ Grav.\  {\bf 22} (2005) 4621
  [arXiv:hep-th/0409229].

\bibitem{Arkani-Hamed:2001ca}
  N.~Arkani-Hamed, A.~G.~Cohen and H.~Georgi,
  Phys.\ Rev.\ Lett.\  {\bf 86} (2001) 4757
  [arXiv:hep-th/0104005].

\bibitem{Hill:2000mu}
  C.~T.~Hill, S.~Pokorski and J.~Wang,
  Phys.\ Rev.\ D {\bf 64} (2001) 105005 
  [arXiv:hep-th/0104035].

\bibitem{Csaki:2003sh}
  C.~Csaki, C.~Grojean, J.~Hubisz, Y.~Shirman and J.~Terning,
  Phys.\ Rev.\ D {\bf 70} (2004) 015012
  [arXiv:hep-ph/0310355].

\bibitem{Carena:2002me}
  M.~Carena, T.~M.~P.~Tait and C.~E.~M.~Wagner,
  Acta Phys.\ Polon.\ B {\bf 33} (2002) 2355
  [arXiv:hep-ph/0207056].

\bibitem{delAguila:2003kd}
  F.~del Aguila, M.~P\'erez-Victoria and J.~Santiago,
  arXiv:hep-ph/0305119.

\bibitem{delAguila:2003gv}
  F.~del Aguila, M.~P\'erez-Victoria and J.~Santiago,
  Acta Phys.\ Polon.\ B {\bf 34} (2003) 5511
  [arXiv:hep-ph/0310353].

\bibitem{Davoudiasl:2002ua}
  H.~Davoudiasl, J.~L.~Hewett and T.~G.~Rizzo,
  Phys.\ Rev.\ D {\bf 68} (2003) 045002
  [arXiv:hep-ph/0212279].

\bibitem{Carena:2002dz}
  M.~Carena, E.~Ponton, T.~M.~P.~Tait and C.~E.~M.~Wagner,
  Phys.\ Rev.\ D {\bf 67} (2003) 096006
  [arXiv:hep-ph/0212307].

\bibitem{Davoudiasl:2003zt}
  H.~Davoudiasl, J.~L.~Hewett and T.~G.~Rizzo,
  JHEP {\bf 0308} (2003) 034
  [arXiv:hep-ph/0305086].

\bibitem{Carena:2004zn}
  M.~Carena, A.~Delgado, E.~Ponton, T.~M.~P.~Tait and C.~E.~M.~Wagner,
  Phys.\ Rev.\ D {\bf 71} (2005) 015010
  [arXiv:hep-ph/0410344].

\bibitem{Scrucca:2003ra}
  C.~A.~Scrucca, M.~Serone and L.~Silvestrini,
  Nucl.\ Phys.\ B {\bf 669} (2003) 128
  [arXiv:hep-ph/0304220].

\bibitem{Barbieri:2003pr}
  R.~Barbieri, A.~Pomarol and R.~Rattazzi,
  Phys.\ Lett.\ B {\bf 591} (2004) 141
  [arXiv:hep-ph/0310285].

\bibitem{Cacciapaglia:2004jz}
  G.~Cacciapaglia, C.~Csaki, C.~Grojean and J.~Terning,
  Phys.\ Rev.\ D {\bf 70} (2004) 075014
  [arXiv:hep-ph/0401160].

\bibitem{Cacciapaglia:2004rb}
  G.~Cacciapaglia, C.~Csaki, C.~Grojean and J.~Terning,
  Phys.\ Rev.\ D {\bf 71} (2005) 035015
  [arXiv:hep-ph/0409126].

\bibitem{Panico:2005ft}
  G.~Panico and M.~Serone,
  JHEP {\bf 0505} (2005) 024
  [arXiv:hep-ph/0502255].

\bibitem{Kolanovic:2003am}
  M.~Kolanovic, M.~Porrati and J.~W.~Rombouts,
  Phys.\ Rev.\ D {\bf 68} (2003) 064018
  [arXiv:hep-th/0304148].

\bibitem{Cheng:2002ab}
  H.~C.~Cheng, K.~T.~Matchev and M.~Schmaltz,
  Phys.\ Rev.\ D {\bf 66} (2002) 056006
  [arXiv:hep-ph/0205314].

\bibitem{DaRold:2003yi}
  L.~Da Rold,
  Phys.\ Rev.\ D {\bf 69} (2004) 105015
  [arXiv:hep-th/0311063].

\bibitem{Contino:2001nj}
  R.~Contino, L.~Pilo, R.~Rattazzi and A.~Strumia,
  JHEP {\bf 0106} (2001) 005
  [arXiv:hep-ph/0103104].

\bibitem{Barvinsky:2005ms}
  A.~O.~Barvinsky and D.~V.~Nesterov,
  arXiv:hep-th/0512291.

\bibitem{Ponton:2005kx}
  E.~Ponton and L.~Wang,
  arXiv:hep-ph/0512304.

\bibitem{Perez-Victoria:2004ef}
  M.~P\'erez-Victoria,
  Acta Phys.\ Polon.\ B {\bf 35} (2004) 2795.

\end{thebibliography}
\end{document}